\documentclass[twocolumn,aps,prd,showpacs,preprintnumbers,floatfix,nofootinbib,superscriptaddress]{revtex4}
\usepackage{graphicx}
\newcommand{\be}{\begin{eqnarray}}
\newcommand{\ee}{\end{eqnarray}}
\begin{document}
%
%\preprint{gr-qc/yymmddd}
%
%
\title{CMB acoustic scale in the entropic-like accelerating universe}
\author{R.~Casadio}
\email{casadio@bo.infn.it}
\affiliation{Dipartimento di Fisica, Universit\`a di Bologna, via Irnerio 46, 40126 Bologna, Italy}
\affiliation{INFN, Sezione di Bologna, Via Irnerio 46, I-40126 Bologna, Italy}
\author{A.~Gruppuso}
\email{gruppuso@iasfbo.inaf.it}
\affiliation{INAF-IASF Bologna, 
Istituto di Astrofisica Spaziale e Fisica Cosmica di Bologna,
Istituto Nazionale di Astrofisica, 
via Gobetti 101, I-40129 Bologna, Italy}
\affiliation{INFN, Sezione di Bologna,
Via Irnerio 46, I-40126 Bologna, Italy}
%
%
%\date{\today} 
%
\begin{abstract} 
We consider generalizations of the entropic accelerating universe recently proposed 
in Ref.~\cite{Easson:2010av,Easson:2010xf} and show that their background equations can be
made equivalent to a model with a dark energy component with constant parameter
of state $w_{X} = -1 + 2\, \gamma /3$, where $\gamma$
is related to the coefficients of the new terms in the Friedman equations.
After discussing all the Friedman equations for an arbitrary $\gamma$,
we show how to recover the standard scalings for dust and radiation.
The acoustic scale $\ell_A$, related to the peak positions in the pattern of the
angular power spectrum of the Cosmic Microwave Background anisotropies,
is also computed and yields the stringent bound $|\gamma|\ll 1$.
We then argue that future data might be able to distinguish this model
from pure $\Lambda$CDM (corresponding to $\gamma=0$).
\end{abstract}
\maketitle
\section{Introduction}
\label{intro}
Modern theoretical cosmology includes an early period of accelerated expansion
named {\em inflation\/}~\cite{inflation}, whose driving force is commonly modeled
using a scalar field (the {\em inflaton\/}) of uncertain nature. 
A similarly accelerated phase is undergoing now~\cite{supernovae} and has led
to conceive the existence of an equally unspecified {\em dark energy\/} component
in the matter content of the Universe \cite{DarkEnergy}.
\par
An alternative scenario has been recently proposed in
Refs.~\cite{Easson:2010av,Easson:2010xf}, based on the idea of entropic gravity
introduced in Ref.~\cite{verlinde}.
In this context, the equations governing the time-evolution of the cosmic scale
factor contain terms proportional to the Hubble function squared $H^2$ and
its time derivative $\dot H$ originating at the boundary of spatial sections of
our universe.
According to the authors of Refs.~\cite{Easson:2010av,Easson:2010xf},
such terms could explain the acceleration occurring both in the early stages
and at present.
Boundary terms, whose nature is well-known in
General Relativity~\cite{Carroll:1997ar},
have indeed been analyzed in various contexts,
for example in Refs.~\cite{otherPaper}.
\par
In this work, we will not analyze how these terms emerge from an
action principle, nor if a unique Lagrangian can be defined at all. 
We shall instead assume general modifications of the form considered in
Refs.~\cite{Easson:2010av,Easson:2010xf} and then try to constrain
their possible effectiveness by comparing the corresponding Cosmic Microwave
Background (CMB) acoustic scale (see, e.g., Refs.~\cite{Page:2003fa,Gruppuso:2005xy})
with the most recent available WMAP data~\cite{Komatsu:2010fb}.
Note that a standard Monte-Carlo Markov Chain analysis 
(usually employed to extract the cosmological parameters
by comparison with available observations) 
is not feasible if the model is unknown at the linear order.
On the contrary, the CMB acoustic scale can be computed directly
from the background equations, and this will allow us to obtain a constraint
for the free parameters of the model by comparing with the most
recent CMB data.  
\par
The paper is organized as follows:
in Section~\ref{mfe} we obtain the complete set of Friedman equations
for a generalization of the entropic models introduced in
Refs.~\cite{Easson:2010av,Easson:2010xf}.
This, in Section~\ref{effDE}, will allow us to regard the model in terms of an effective
dark energy contribution depending on one parameter $\gamma$.
In particular, bounds on $\gamma$ will be obtained by comparing
the CMB acoustic scale with the 7yr~WMAP data in Section~\ref{acousticscale},
after computing the deceleration parameter in Section~\ref{decpar}.
Conclusions will be drawn in Section~\ref{conclusions}.
\section{Modified Friedman equations}
\label{mfe}
In the flat Friedman-Robertson-Walker metric 
\be
ds^2 = -dt^2 + a^2(t)\, d \vec x\cdot d\vec x
\ ,
\ee
with the scale factor $a(t)$ normalized so that $a(t_{\rm now})=1$,
the model of universe considered in Refs.~\cite{Easson:2010av,Easson:2010xf}
features a Friedman equation given by
\be
\frac{\ddot a}{a}
=
- \frac{4\, \pi\, \tilde G}{ 3}\, \sum_i \left( \tilde\rho_i + 3\, \tilde p_i \right) + C_H\, H^2 + C_{\dot H}\, \dot H
\ ,
\label{eqFriedman}
\ee
where $H = \dot a / a\equiv a^{-1}\,da/dt$, $\tilde G$ is the ``bare'' Newton constant,
$\tilde \rho_i$ and $\tilde p_i$ are the ``bare'' energy density and pressure of
the $i$-th fluid filling the universe, while $C_H$ and $C_{\dot H}$ are constants 
coming from the boundary terms.
As already stated in the Introduction, we take such terms as given and do
not derive them from the Einstein-Hilbert action on a manifold with boundaries.
Instead, we shall determine the full set of cosmological (and continuity) equations
consistent with Eq.~(\ref{eqFriedman}) without {\em a priori\/} fixing
$C_H$ and $C_{\dot H}$.
\par 
We first note that Eq.~(\ref{eqFriedman}) can be rewritten as
\be
\dot H + \gamma \,H^2
= 
- \frac{4\, \pi\, G}{ 3}\, \sum_i \left( \tilde \rho_i + 3\,\tilde p_i \right)
\, ,
\label{eqFriedman2}
\ee
where
\be
\gamma=\frac{1 - C_H}{1- C_{\dot H}}
\ ,
\label{ab}
\ee
and we rescaled~\footnote{This rescaling~\cite{sorbo} and Eq.~(\ref{ab}) are meaningful only
if $C_{\dot H} \neq 1$, a condition we assume throughout the paper.
If $C_{\dot H} =1$, Eq.~(\ref{eqFriedman2}) does not contain $\dot H$ and is
therefore not an equation of motion but a constraint.}
\be
G=\tilde G/(1- C_{\dot H})
\ .
\label{newton}
\ee
Noting that $d/dt=(a\,H)\,d/da$ and assuming
\be
\tilde \rho_i = \tilde\rho_i^{(0)} \, a^{-k_i}
\ ,
\quad
\tilde p_i=w_i\,\tilde \rho_i
\ ,
\label{w}
\ee
with $ \tilde\rho_i^{(0)}$ and $w_i$ constant,
Eq.~(\ref{eqFriedman2}) can be integrated exactly and yields
\be
H^2 = \frac{8\, \pi\, G}{ 3}\, \left( \sum_i \, c_i \,\tilde\rho_i + \frac{C}{a^{2\,\gamma}} \right)
\ ,
\label{eqFriedman3}
\ee
where the coefficients
\be
c_i=\frac{1+3\,w_i}{k_i-2\,\gamma}
\label{ci}
\ee
are well-defined only for $k_i\not=2\,\gamma$
and $C$ is a constant of integration.
Further, on deriving Eq.~(\ref{eqFriedman3}) with respect to time and using
Eqs.~(\ref{eqFriedman2}) and~(\ref{ci}), we obtain the continuity equation
\be
\dot{\tilde \rho}_i + \frac{H}{c_i} \left[ \left(2 \, \gamma \, c_i + 1\right)\tilde\rho _i+ 3 \,\tilde p_i \right] 
=\dot{\tilde\rho}_i+H\,k_i\,\tilde\rho_i
=0
\ ,
\label{coneq}
\ee
which is identically satisfied for the fluids~(\ref{w}).
For example, for dust we have $w_{\rm dust} =0$ and requiring $k_{\rm dust}=3$
yields $c_{\rm dust}=1/(3 -2\, \gamma)$.
Likewise, radiation has $w_{\rm rad}=1/3$ and requiring $k_{\rm rad}=4$ results in
$c_{\rm rad}=1/(2 - \gamma)$.
Specifying these parameters and assuming that the matter content of the universe
is a mixture of dust and radiation, the Friedman equations~(\ref{eqFriedman2}) and
(\ref{eqFriedman3}) can be rewritten as (see also Appendix \ref{appendice})
\begin{eqnarray}
\frac{\ddot a}{a}
&\!=\!&
- \frac{4\, \pi\, G}{ 3} \left[
\frac{\rho_{\rm dust}^{(0)}}{a^3}
+  2 \, \frac{\rho_{\rm rad}^{(0)}}{a^4}
-2\,(1-\gamma) \, \frac{C}{a^{2  \gamma}}
\right]
\label{EqFriedman_ddota_eff}
\\
H^2
&\!=\!&
\frac{8\, \pi\, G}{ 3} \left[
\frac{\rho_{\rm dust}^{(0) }}{a^3}
+  \frac{\rho_{\rm rad}^{(0) }}{a^4}
+ \frac{C}{a^{2  \gamma}}
\right] 
\label{EqFriedman_dota_eff}
\ ,
\end{eqnarray}
where $\rho_{\rm dust}^{(0) } = c_{\rm dust }\,\tilde\rho_{\rm dust}^{(0)}$
and $\rho_{\rm rad}^{(0) } = c_{\rm rad}\,\tilde\rho_{\rm rad}^{(0)}$ are
the present matter and radiation densities involved in observations.
Note that the (bare) densities $\tilde\rho_{\rm rad}$ and $\tilde\rho_{\rm dust}$
and the corresponding constant and dimensionless coefficients $c_{\rm rad}$
and $c_{\rm dust}$ are not observable separately, since only their products
appear in the equations (as we remark in Appendix~\ref{appendice}).
\par
Eqs.~(\ref{EqFriedman_ddota_eff}) and (\ref{EqFriedman_dota_eff})
are precisely the standard Friedman equations for a universe filled with dust
and radiation that scale in the usual way, namely
\be
\rho_{\rm dust}=\rho_{\rm dust}^{(0)}/a^3
\ ,
\qquad
\rho_{\rm rad}=\rho_{\rm rad}^{(0)}/a^4
\ ,
\label{newrho}
\ee
corrected by terms proportional to $C$.
Note also that, in the limit $\gamma \to 1$, we recover 
the standard cosmological equations (with no dark energy component!) 
with the $C$-term playing the role of an effective curvature contribution.
However, for $\gamma\not=1$, there is a region where the corrections
can be interpreted as an effective dark energy component if $C >0$.
This is the case we consider in the following, with 
$\rho_{\rm dust}^{(0)}$ and $\rho_{\rm rad}^{(0)}$
equal to the present dust and radiation densities.
For example, $\rho_{\rm rad}^{(0)}$ is the present energy density of
the black-body radiation with temperature $T=2.725\,$K (multiplied by the
contribution from neutrinos).
\section{Effective Dark Energy}
\label{effDE}
The next step is to find whether there exist values of $\gamma$ 
corresponding to an accelerating universe, i.e.,~such that $\ddot a >0$.
This can be understood analyzing Eq.~(\ref{EqFriedman_ddota_eff}).
The effective dark energy term proportional to $C$ will then drive
the present acceleration of the universe if it dominates in the r.h.s.~of
Eq.~(\ref{EqFriedman_ddota_eff}) at recent times.
We therefore require that $\gamma< 3/2$.
Moreover, since we also assume $C>0$, $\ddot a>0$ implies
\be
\gamma <1
\ .
\label{b<a}
\ee
Hence, when Eq.~(\ref{b<a}) is satisfied, the $C$-term
mimics the behavior of a dark energy fluid
[see Eq.~(\ref{EqFriedman_dota_eff})]
with constant parameter of state $w_X = -1 + 2\, \gamma / 3$.
\subsection{Deceleration parameter}
\label{decpar}
\begin{figure}
\includegraphics[width=8.0cm]{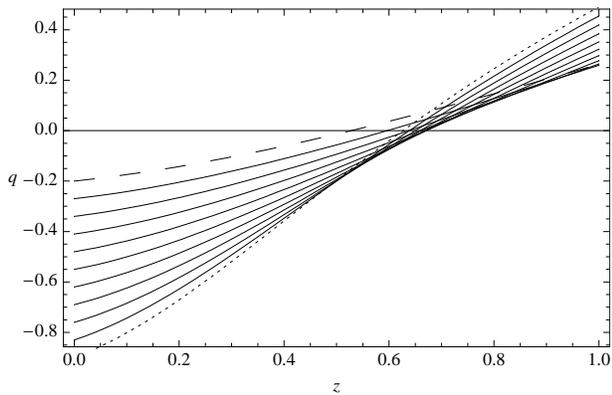}
\caption{$q$ vs $z$ for $\gamma$ from $-0.5$ to $0.5$ with step equal to $0.1$.
Dotted line is for $\gamma=-0.5$, dashed line for $\gamma=0.5$ and
solid lines for values in between.}
\label{FigDue}
\end{figure}

The deceleration parameter is defined as
\be
q = - \frac{\ddot a}{a\, H^2}
\ .
\label{defq}
\ee
Plugging Eqs.~(\ref{EqFriedman_ddota_eff}) and~(\ref{EqFriedman_dota_eff})
into the above definition and neglecting radiation~\footnote{Of course, this
approximation is valid for recent cosmological times (like in Fig.~\ref{FigDue}),
when the transition from matter to dark energy dominated epochs was taking place
and the contribution of radiation was subleading.}
yields
\be
q =
\frac{1}{2 \, a }
\left( \frac{\Omega_C\,a^{-3} - 2 \, (1-\gamma)\, \Omega_{\Lambda}\,a^{-2 \gamma}}
{\Omega_C\,a^{-3} +  \Omega_{\Lambda}\,a^{-2\gamma}}\right)
\ ,
\ee
where $\Omega_{\rm C} = \rho_{\rm dust}^{(0)} / \rho_{\rm c}$, 
$\Omega_{\Lambda} = C  / \rho_{\rm c}$ and $\rho_{\rm c} \equiv 3\, H_0^2/ (8\, \pi\, G)$
with $H_0$ the present value of the Hubble function.
\par
In Fig.~\ref{FigDue}, we show $q$ as a function of the redshift $z$ for various values of
$\gamma$ from $-0.5$ to $0.5$ in steps of $0.1$.
Note that, on specializing $q$ at present time, we find
\be
q = \frac{1}{2}\, \Omega_C - (1-\gamma)\, \Omega_{\Lambda}
\ ,
\ee
which turns out to be the standard expression for the $\Lambda$CDM model when
$\gamma =0$~\cite{Sahni:2002fz,Visser:2003vq,Gruppuso:2005xy}.
\subsection{CMB acoustic scale}
\label{acousticscale}
The characteristic angular scale $\theta_A$ of the peaks of the angular power
spectrum in CMB anisotropies is defined as~\cite{Page:2003fa}
\be
\theta_A = {\frac{r_{\rm s}(z_{\rm dec})}{r(z_{\rm dec})}} = \frac{\pi}{\ell_A}
\ ,
\label{acousticangularscale}
\ee
where $r_{\rm s}(z_{\rm dec})$ is the comoving size of the sound horizon at decoupling,
$r(z_{\rm dec})$ the comoving distance at decoupling and $\ell _A$ the multipole associated
with the angular scale $\theta_A$, also called the {\em acoustic scale\/}.
Let us recall that $\ell_A$ is not exactly the scale of the first peak.
In general, the position of the $m$-th peak is given by $ \ell_m = \left( m - \phi_m \right) \ell_A $
where $\phi_m$ is a phase that depends on other cosmological parameters~\cite{Page:2003fa}.
\par
In order to make explicit the dependence of $\ell_A$ on the cosmological parameters,
we now consider separately numerator and denominator of  Eq.~(\ref{acousticangularscale}).
The comoving size of the sound horizon at decoupling can be written as~\cite{huandsugiyama}
\be
r_{s}(z_{\rm dec})
&\!\!=\!\!&
\frac{4}{3\, H_0}\, \sqrt{\frac{\Omega_{\gamma}}{\Omega_C\, \Omega_b}}
\nonumber
\\ 
&&
\times
\ln \left[
\frac{\sqrt{1+ R_{\rm dec}} + \sqrt{R_{\rm dec}+R_{\rm eq}}}{1+\sqrt{R_{\rm eq}}}
\right]
\ ,
\label{rscomputed}
\ee
with $R(z)= 3 (\Omega_b/ (4 \Omega_{\gamma})) / (1+z)$ and where $\Omega_b$
and $\Omega_{\gamma}$ are the present density ratios for baryons
and photons respectively [note the index $\gamma$ in $\Omega_{\gamma}$
must not to be confused with the parameter $\gamma$ defined in Eq.~(\ref{ab})].
Moreover, the label ``dec'' stands for ``computed at decoupling'', while ``eq''
stands for ``computed at equivalence'' (between radiation and matter).
By definition the comoving distance at decoupling reads
\be
r (z_{\rm dec}) = \int_0^{z_{\rm dec}} \frac{d z'}{H(z')}
\ ,
\label{cdz}
\ee
where $H(z)$ is given by Eq.~(\ref{EqFriedman_dota_eff}) and can be recast as
\be
H(z)
&\!\!=\!\!&
H_0 \left[ (1+z)^3 \, \Omega_C + (1+z)^4 \, \Omega_{\rm rad} +
\right.
\nonumber
\\ 
&&
\left.
+ (1+z)^{2 \gamma} \, \Omega_{\Lambda} \right]^{1/2}
\ ,
\label{Hznew}
\ee
where $ \Omega_{\rm rad} = \rho_{\rm rad}^{(0)} / \rho_{\rm c}$.
We can therefore write the acoustic scale $\ell_A$ as
\begin{widetext}
\be
\ell_A
=
\frac{3\, \pi}{4}\, \sqrt{\frac{\Omega_b}{\Omega_\gamma}}\,
\frac{ \int_0^{z_{\rm dec}} dz
\left[ (1+z)^3 + (1+z)^4 (\Omega_{\rm rad}/\Omega_C)
+ (1 +z)^{2 \gamma} (\Omega_{\Lambda}/\Omega_C) \right]^{-1/2}}
{\ln \left[ \sqrt{1+ R_{\rm dec}} + 
\sqrt{R_{\rm dec}+R_{\rm eq}} \right]
- \ln\left[ {1+\sqrt{R_{\rm eq}}} \right]}
\, .
\label{la}
\ee
\end{widetext}
Let us remark that Eq.~(\ref{la}) was obtained by neglecting $\Omega_{\Lambda}$
in  $r_{\rm s}(z_{\rm dec})$ (the comoving size of the sound horizon at decoupling).
However, it was shown in Ref.~\cite{Gruppuso:2005xy} that this approximation
at most leads to $10^{-5}\,\%$ error, much smaller than the precision of our
result below [see Eq.~(\ref{rangepergamma})].
\par
\begin{figure}
\includegraphics[width=8.0cm]{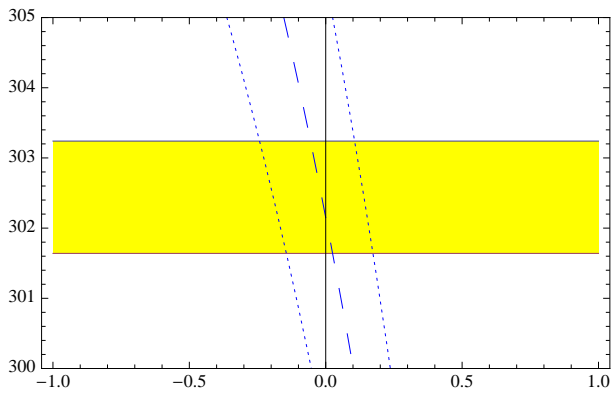}
\caption{Acoustic scale for entropic universe as function of $\gamma$.
Horizontal lines represent 1-$\sigma$ WMAP measurement (colored region displays
1-$\sigma$ contour).
Long dashed line is for $\ell_A$ computed at best fit of $z_{\rm eq}=3196$.
Short dashed lines are for $\ell_A$ computed at 1-$\sigma$ level of
$z_{\rm eq}=3196^{+ 134}_{-133}$.}
\label{FigTwo}
\end{figure}

Eq.~(\ref{la}) can now be used to constrain the models under study by
comparing with the value obtained from the recent 7yr~WMAP
data~\cite{Komatsu:2010fb}~\footnote{http://lambda.gsfc.nasa.gov/
\label{foot1}}
\be
\ell_A^{\rm WMAP} = 302.44 \pm 0.8
\ .
\label{WMAPvalue}
\ee
Note that our choices for $\rho_{\rm dust}^{(0)}$ and $\rho_{\rm rad}^{(0)}$
were made in order to minimize deviations from the $\Lambda$CDM model.
In fact, departures of the background equations~(\ref{EqFriedman_ddota_eff})
and (\ref{EqFriedman_dota_eff}) from $\Lambda$CDM are completely parameterized 
by the single parameter $\gamma$ and we are therefore allowed to estimate
Eq.~(\ref{la}) with the values of the other parameters that best fit WMAP data.
We insert in Eq.~(\ref{la}) the 7yr~WMAP best fit values~\cite{Komatsu:2010fb}
$\Omega_b = 0.0449$,
$\Omega_{\gamma}=4.89 \times 10^{-5}$, $z_{\rm dec}=1088.2$, $z_{\rm eq}=3196$,
$\Omega_C=0.266$, $\Omega_{\Lambda}=0.734$ and
$\Omega_{\rm rad}=1.69 \, \Omega_{\gamma}$.
In Fig.~\ref{FigTwo}, we show the acoustic scale (long and short dashed lines) versus
$\gamma$, along with the 1-$\sigma$ levels of the WMAP measurement
(solid horizontal lines).
We also display the dependence of the acoustic scale on $z_{\rm eq}$:
the long dashed line stands for the 7yr~WMAP best fit ($z_{\rm eq}=3196$) whereas 
the short dashed lines stand for its 1-$\sigma$ values, $z_{\rm eq}=3196^{+ 134}_{-133}$.
\par
As Fig.~\ref{FigTwo} shows clearly, the parameter $\gamma$ must be very close to $0$
in order to have consistency with the WMAP observations.
More precisely, from Eq.~(\ref{la}) computed at the best fit values of the
7yr~WMAP parameters~\footnote{This means taking into
account only the long dashed line in Fig.~\ref{FigTwo}.},
we obtain
\be
\gamma = -0.02 \pm 0.04
\ .
\label{rangepergamma}
\ee
This result is consistent with Eq.~(\ref{b<a}) and
implies $-1.040 < w_X < $ -0.986,
so that the added contributions must closely mimic a
cosmological constant.
Further, from Eq.~(\ref{ab}), this implies that
\be
|1-C_H|
\ll
|1-C_{\dot H}|
\label{cCC}
\ee
in Eq.~(\ref{eqFriedman}). 
For example, if $0<C_H, C_{\dot H}<1$, then the strong inequality~(\ref{cCC})
is satisfied for $C_H \approx 1$. 
\section{Conclusions}
\label{conclusions}
In the present paper, we have shown how it is possible to recover standard
background scalings for radiation and matter and standard effective cosmological
equations [see Eqs.~(\ref{EqFriedman_ddota_eff}) and (\ref{EqFriedman_dota_eff})] 
when the Friedman equation for $\ddot a$ is modified by adding
terms proportional to $H^2$ and $\dot H$ like in Eq.~(\ref{eqFriedman}).
An example that requires such a modification is given by the
entropic accelerating universe of Refs.~\cite{Easson:2010av,Easson:2010xf},
although our considerations are more general.
Moreover we have shown how to obtain the recent cosmological acceleration
within the considered model, without adding a dark energy fluid.
We note that for the range of parameters considered here,
the model under analysis does not modify the evolution of the universe when
it was matter or radiation dominated.
Therefore, none of the standard cosmological constraints coming from such early
epochs, as for instance the Big Bang Nucleosynthesis (BBN), are affected.
\par
Specifically, we have shown that the parameter space admits a region
(i.e., $\gamma < 1$) where the universe accelerates at recent cosmological times
(i.e., $z \sim 0.5$).
In fact, the additional terms mimic the behavior of a fluid with a constant
parameter of state $w_X = -1 + 2\, \gamma /3$.
This has been studied by computing the deceleration parameter $q$
[see Fig.~\ref{FigDue} and Section~\ref{decpar}] and stringent constraints
have been obtained comparing the CMB acoustic scale $\ell_A$
with the WMAP~7yr release data. 
Note that a standard Monte-Carlo Markov Chain analysis 
(usually employed to extract the cosmological parameters
by comparison with available observations) is not feasible if the model
is not known at linear order~\footnote{In fact, as far as we know, no Lagrangian
is known for these models. 
However, as we stated in the Introduction, we are not debating 
the theoretical ground the ``entropic-like'' proposal is based on,
but are rather interested in which constraints we can provide
for such class of models from what we have at hand, i.e.~the
background equations.}.
On the contrary, the CMB acoustic scale can be computed directly
from the background equations, and this has allowed us to obtain a constraint
for the free parameters of the model by comparing with the most
recent CMB data.
This comparison has told us that $|\gamma|\ll 1$ (so that $w_X\simeq -1$) and the
coefficients of $\dot H$ and of $H^2$ in Eq.~(\ref{eqFriedman}) must therefore
satisfy Eq.~(\ref{cCC}) for the model to be phenomenologically viable.
In particular, the entropic accelerating universe corresponds to a specific choice
of the constants $C_H$ and $C_{\dot H}$, that
is $\gamma_{\rm I} = 0$ and $\gamma_{\rm II}=0.68$ for the two cases
explicitly mentioned in Ref.~\cite{Easson:2010av}.
The latter is at odd with the constraint~(\ref{rangepergamma}),
whereas the former is consistent.
\par
Future CMB observations coming from the {\sc Planck} satellite~\footnote{Planck
(http://www.esa.int/Planck) is a project of the European Space Agency, ESA.}
are expected to improve the error on the acoustic scale by about
one order of magnitude~\cite{Colombo:2008ta}.
The same improvement is therefore expected for the estimate of the parameter
$\gamma$, which, in principle, should allow us to distinguish these
models from the pure $\Lambda$CDM model with $\gamma=0$.
We finally mention that our findings are in agreement with those of
Ref.~\cite{koivisto}.
\begin{acknowledgments}
We thank F.~Finelli for fruitful discussions.
R.C.~would like to thank D.~Easson and R.~Woodard.
A.G.~also thanks L.~Sorbo for comments on the manuscript
and L.~Colombo for some clarification on Ref.~\cite{Colombo:2008ta}.
We acknowledge the use of the Legacy Archive for Microwave Background
Data Analysis (LAMBDA).
Support for LAMBDA is provided by the NASA Office of Space Science.
A.G.~acknowledges support by ASI through ASI/INAF agreement
I/072/09/0 for the Planck LFI activity of Phase E2.
\end{acknowledgments}
\appendix
\section{Derivation of Friedman equations}
\label{appendice}
From $\dot H = {\ddot a}/ a - H^2$, we can rewrite Eq.~(\ref{eqFriedman2}) as
\be
 \frac{\ddot a}{a} + (\gamma -1) \,H^2
= 
- \frac{4\, \pi\, G}{ 3}\, \sum_i \left(\tilde \rho_i + 3\,\tilde p_i \right)
\ ,
\label{step1}
\ee
in which we note the use of $G$ (instead of the bare $\tilde G$).
We then replace $H^2$ from Eq.~(\ref{eqFriedman3}) into Eq.~(\ref{step1}) and specifying
only two fluids (dust and radiation, with equations of state $\tilde p_{\rm dust}=0$ and
$\tilde p_{\rm rad} = \tilde \rho_{\rm rad}/3$) we obtain
\be
\!\!\!\!\!\!\!\!\!\!
\frac{\ddot a}{a} 
&\!\!=\!\!& 
- \frac{4\, \pi\, G}{ 3}
\left(\tilde\rho_{\rm dust} + 2\,\tilde \rho_{\rm rad} + 2 \, c_{\rm dust} \, \tilde\rho_{\rm dust} \, (\gamma -1)
\phantom{\frac{a}{b}}
\right.
\nonumber 
\\
\!\!\!\!\!\!\!\!\!\!
&&
\phantom{- \frac{4\, \pi\, G}{ 3}}
\left.
+
2 \, c_{\rm rad} \, \tilde\rho_{\rm rad} \, (\gamma -1) + 2\, (\gamma-1)\, \frac{C}{a^{2 \gamma}}\right)
.
\label{step2}
\ee
From the definition of $c_{\rm dust}$ and  $c_{\rm rad}$, it is easy to show that
\be
2 \, c_{\rm dust} \, (\gamma-1) +1 = c_{\rm dust}
\ee
and
\be
2 \, c_{\rm rad} \, (\gamma-1) +2 = 2 \, c_{\rm rad}
\ .
\ee
Therefore, Eq.~(\ref{step2}) is equivalent to 
\be
\frac{\ddot a}{a} 
&\!\!=\!\!&
- \frac{4\, \pi\, G}{ 3}
\left(c_{\rm dust}\,\tilde \rho_{\rm dust} + 2 \, c_{\rm rad}\, \tilde\rho_{\rm rad} \phantom{\frac{a}{b}} 
\right.
\nonumber
\\
&&
\phantom{- \frac{4\, \pi\, G}{ 3}}
\left.
+
2 \,(\gamma-1)\, \frac{C}{a^{2 \gamma}} \right)
\ ,
\label{step3}
\ee
which is exactly Eq.~(\ref{EqFriedman_ddota_eff}) with the densities of Eq.~(\ref{newrho}),
namely
\be
\rho_{\rm rad}=c_{\rm rad}\, \tilde\rho_{\rm rad}
\ ,
\qquad
\rho_{\rm dust}=c_{\rm dust}\, \tilde\rho_{\rm dust}
\ .
\ee
Note that the constant and dimensionless coefficients $c_{\rm rad}$ and $c_{\rm dust}$
are not observable, since they never appear without multiplying the corresponding (bare)
densities, and the above rescaling simply reflects the choice of standard units for the
densities [as well as Eq.~(\ref{newton}) is for the Newton constant]. 
\par
Finally, Eq.~(\ref{eqFriedman3}) becomes Eq.~(\ref{EqFriedman_dota_eff}) using
the same redefinitions (choice of units) for the densities.

\begin{thebibliography}{99}
\bibitem{inflation}
A.H.~Guth,
Phys.\ Rev.\  D {\bf 23}, 347 (1981);
%%CITATION = PHRVA,D23,347;%%
%
A.D.~Linde,
Phys.\ Lett.\  B {\bf 108}, 389 (1982);
%%CITATION = PHLTA,B108,389;%%
A.J.~Albrecht and P.J.~Steinhardt,
Phys.\ Rev.\ Lett.\  {\bf 48}, 1220 (1982);
%%CITATION = PRLTA,48,1220;%%
A.A.~Starobinsky, Phys.\ Lett.\ B 91, 99 (1980).
%
\bibitem{supernovae}
S.~Perlmutter {\it et al.},
Astrophys.\ J.\  {\bf 517}, 565 (1999);
%%CITATION = ASJOA,517,565;%%
A.G.~Riess {\it et al.},
Astron.\ J.\  {\bf 116}, 1009 (1998).
%%CITATION = ANJOA,116,1009;%%
%
\bibitem{DarkEnergy}
%\cite{Peebles:2002gy}
%\bibitem{Peebles:2002gy}
  P.J.E.~Peebles and B.~Ratra,
  %``The cosmological constant and dark energy,''
  Rev.\ Mod.\ Phys.\  {\bf 75} (2003) 559;
  %%CITATION = RMPHA,75,559;%%
%
%\cite{Copeland:2006wr}
%\bibitem{Copeland:2006wr}
  E.J.~Copeland, M.~Sami and S.~Tsujikawa,
  %``Dynamics of dark energy,''
  Int.\ J.\ Mod.\ Phys.\  D {\bf 15} (2006) 1753.
  %%CITATION = IMPAE,D15,1753;%%
%
%\cite{Easson:2010av}
\bibitem{Easson:2010av}
  D.A.~Easson, P.H.~Frampton and G.F.~Smoot,
% ``Entropic Accelerating Universe,''
  Phys.\ Lett.\  B {\bf 696}, 273 (2011).
  %%CITATION = PHLTA,B696,273;%%
%
%\cite{Easson:2010xf}
\bibitem{Easson:2010xf}
  D.A.~Easson, P.H.~Frampton and G.F.~Smoot,
  ``Entropic Inflation,''
  arXiv:1003.1528 [hep-th].
  %%CITATION = ARXIV:1003.1528;%%
%
\bibitem{verlinde}
E.P.~Verlinde,
``On the Origin of Gravity and the Laws of Newton,''
arXiv:1001.0785 [hep-th].
%%CITATION = ARXIV:1001.0785;%%
%
%\cite{Carroll:1997ar}
\bibitem{Carroll:1997ar}
S.M.~Carroll,
  ``Lecture notes on general relativity,''
  arXiv:gr-qc/9712019;
  %%CITATION = GR-QC/9712019;%%
  %\cite{Casadio:2001ff}
%\bibitem{Casadio:2001ff}
R.~Casadio and A.~Gruppuso,
%``On boundary terms and conformal transformations in curved space-times,''
Int.\ J.\ Mod.\ Phys.\  D {\bf 11} (2002) 703.
%%CITATION = IMPAE,D11,703;%%  
%
\bibitem{otherPaper}
%\cite{ChangYoung:2010rz}
%\bibitem{ChangYoung:2010rz}
  E.~Chang-Young, M.~Eune, K.~Kimm and D.~Lee,
  %``Surface gravity and Hawking temperature from entropic force viewpoint,''
  Mod.\ Phys.\ Lett.\  A {\bf 25}, 2825 (2010).
  %[arXiv:1003.2049 [gr-qc]].
  %%CITATION = MPLAE,A25,2825;%%
%
%\cite{Lee:2010ew}
%\bibitem{Lee:2010ew}
  J.W.~Lee,
  ``Zero Cosmological Constant and Nonzero Dark Energy from Holographic Principle,''
  arXiv:1003.1878 [hep-th];
  %%CITATION = ARXIV:1003.1878;%%
%
%\cite{He:2010yf}
%\bibitem{He:2010yf}
  X.G.~He and B.Q.~Ma,
  %``Black Holes and Photons with Entropic Force,''
  Chin.\ Phys.\ Lett.\  {\bf 27}, 070402 (2010);
  %[arXiv:1003.1625 [hep-th]].
  %%CITATION = CPLEE,27,070402;%%
%
%\cite{Danielsson:2010uy}
%\bibitem{Danielsson:2010uy}
  U.H.~Danielsson,
  ``Entropic dark energy and sourced Friedmann equations,''
  arXiv:1003.0668 [hep-th];
  %%CITATION = ARXIV:1003.0668;%%
%
%\cite{Lee:2010fg}
%\bibitem{Lee:2010fg}
  J.W.~Lee, H.C.~Kim and J.~Lee,
 ``Gravity as Quantum Entanglement Force,''
  arXiv:1002.4568 [hep-th];
  %%CITATION = ARXIV:1002.4568;%%
  %\bibitem{Wei:2010wwa}
  S.W.~Wei, Y.X.~Liu and Y.Q.~Wang,
  ``Friedmann equation of FRW universe in deformed Horava-Lifshitz gravity from
  entropic force,''
  arXiv:1001.5238 [hep-th];
  %%CITATION = ARXIV:1001.5238;%%
  %\bibitem{Liu:2010na}
  Y.X.~Liu, Y.Q.~Wang and S.W.~Wei,
  %``A Note on Temperature and Energy of 4-dimensional Black Holes from Entropic
  %Force,''
  Class.\ Quant.\ Grav.\  {\bf 27}, 185002 (2010);
  %[arXiv:1002.1062 [hep-th]];
  %%CITATION = CQGRD,27,185002;%%;
  %\bibitem{Modesto:2010rm}
  L.~Modesto and A.~Randono,
  ``Entropic corrections to Newton's law,''
  arXiv:1003.1998 [hep-th].
  %%CITATION = ARXIV:1003.1998;%%
%
%
%\cite{Page:2003fa}
\bibitem{Page:2003fa}
  L.~Page {\it et al.}  [WMAP Collaboration],
  %``First Year Wilkinson Microwave Anisotropy Probe (WMAP) Observations:
  %Interpretation of the TT and TE Angular Power Spectrum Peaks,''
  Astrophys.\ J.\ Suppl.\  {\bf 148}, 233 (2003).
  %[arXiv:astro-ph/0302220].
  %%CITATION = APJSA,148,233;%%
%\cite{Gruppuso:2005xy}
%
\bibitem{huandsugiyama}
W.~Hu and N.~Sugiyama,
Astrophys.J. {\bf 444}, 489 (1995).
%
\bibitem{Gruppuso:2005xy}
  A.~Gruppuso and F.~Finelli,
  %``Analytic Results for a Flat Universe Dominated by Dust and Dark Energy,''
  Phys.\ Rev.\  D {\bf 73},  023512 (2006).
  %[arXiv:astro-ph/0512641].
  %%CITATION = PHRVA,D73,023512;%%
%
%\cite{Komatsu:2010fb}
\bibitem{Komatsu:2010fb}
  E.~Komatsu {\it et al.},
  %``Seven-Year Wilkinson Microwave Anisotropy Probe (WMAP) Observations:
  %Cosmological Interpretation,''
  Astrophys.\ J.\ Suppl.\  {\bf 192}, 18 (2011).
  %[arXiv:1001.4538 [astro-ph.CO]].
  %%CITATION = APJSA,192,18;%%
%  
%\cite{Sahni:2002fz}
\bibitem{Sahni:2002fz}
  V.~Sahni, T.~D.~Saini, A.~A.~Starobinsky and U.~Alam,
  %``Statefinder -- a new geometrical diagnostic of dark energy,''
  JETP Lett.\  {\bf 77}, 201 (2003)
  [Pisma Zh.\ Eksp.\ Teor.\ Fiz.\  {\bf 77}, 249 (2003)].
 % [arXiv:astro-ph/0201498].
  %%CITATION = ZFPRA,77,249;%%
%  
%\cite{Visser:2003vq}
\bibitem{Visser:2003vq}
  M.~Visser,
  %``Jerk and the cosmological equation of state,''
  Class.\ Quant.\ Grav.\  {\bf 21}, 2603 (2004).
  %[arXiv:gr-qc/0309109].
  %%CITATION = CQGRD,21,2603;%%
%
%\cite{Colombo:2008ta}
\bibitem{Colombo:2008ta}
  L.P.L.~Colombo, E.~Pierpaoli and J.R.~Pritchard,
 % ``Cosmological parameters after WMAP5: forecasts for Planck and future galaxy
  %surveys,''
  Mon.\ Not.\ Roy.\ Astron.\ Soc.\  {\bf 398}, 1621 (2009).
  %[arXiv:0811.2622 [astro-ph]].
  %%CITATION = MNRAA,398,1621;%%
%
\bibitem{koivisto}
  T.S.~Koivisto, D.F.~Mota and M.~Zumalacarregui,
  %``Constraining Entropic Cosmology,''
  JCAP {\bf 1102}, 027 (2011).
 % [arXiv:1011.2226 [astro-ph.CO]].
  %%CITATION = JCAPA,1102,027;%%
%
\bibitem{sorbo}
We thank L.~Sorbo for this observation.
%
\end{thebibliography}
\end{document}